\documentclass[aps,prd,superscriptaddress,onecolumn,floatfix,nofootinbib]{revtex4}
\usepackage{amsmath,graphicx,tabularx,url,multirow,hyperref}

\begin{document}

\def\tablename{Table}
\def\figurename{Figure}

\newcommand{\heptop}{{\sc HEPTopTagger \,}}
\newcommand{\eps}{\varepsilon}

\newcommand\one{\leavevmode\hbox{\small1\normalsize\kern-.33em1}}
\newcommand{\Mpl}{M_\mathrm{Pl}}
\newcommand{\p}{\partial}
\newcommand{\lag}{\mathcal{L}}
\newcommand{\qqquad}{\qquad \qquad}
\newcommand{\qqqquad}{\qquad \qquad \qquad}

\newcommand{\qb}{\bar{q}}
\newcommand{\matx}{|\mathcal{M}|^2}
\newcommand{\really}{\stackrel{!}{=}}
\newcommand{\msbar}{\overline{\text{MS}}}
\newcommand{\qns}{f_q^\text{NS}}
\newcommand{\lqcd}{\Lambda_\text{QCD}}
\newcommand{\met}{\slashchar{p}_T}
\newcommand{\pmiss}{\slashchar{\vec{p}}_T}

\newcommand{\st}[1]{\tilde{t}_{#1}}
\newcommand{\stb}[1]{\tilde{t}_{#1}^*}
\newcommand{\nz}[1]{\tilde{\chi}_{#1}^0}
\newcommand{\cp}[1]{\tilde{\chi}_{#1}^+}
\newcommand{\cm}[1]{\tilde{\chi}_{#1}^-}

\providecommand{\mg}{m_{\tilde{g}}}
\providecommand{\mst}{m_{\tilde{t}}}
\newcommand{\msn}[1]{m_{\tilde{\nu}_{#1}}}
\newcommand{\mch}[1]{m_{\tilde{\chi}^+_{#1}}}
\newcommand{\mne}[1]{m_{\tilde{\chi}^0_{#1}}}
\newcommand{\msb}[1]{m_{\tilde{b}_{#1}}}

\newcommand{\mev}{{\ensuremath\rm MeV}}
\newcommand{\gev}{{\ensuremath\rm GeV}}
\newcommand{\tev}{{\ensuremath\rm TeV}}
\newcommand{\fb}{{\ensuremath\rm fb}}
\newcommand{\ab}{{\ensuremath\rm ab}}
\newcommand{\pb}{{\ensuremath\rm pb}}
\newcommand{\sign}{{\ensuremath\rm sign}}
\newcommand{\ifb}{{\ensuremath\rm fb^{-1}}}

\def\slashchar#1{\setbox0=\hbox{$#1$}           
   \dimen0=\wd0                                 
   \setbox1=\hbox{/} \dimen1=\wd1               
   \ifdim\dimen0>\dimen1                        
      \rlap{\hbox to \dimen0{\hfil/\hfil}}      
      #1                                        
   \else                                        
      \rlap{\hbox to \dimen1{\hfil$#1$\hfil}}   
      /                                         
   \fi}
\newcommand{\dslash}{\slashchar{\partial}}
\newcommand{\Dslash}{\slashchar{D}}

\title{Benchmarking an Even Better HEPTopTagger}

\author{Christoph Anders}
\affiliation{Physikalisches Institut, 
             Universit\"at Heidelberg, Germany}

\author{Catherine Bernaciak}
\affiliation{Institut f\"ur Theoretische Physik, 
             Universit\"at Heidelberg, Germany}

\author{Gregor Kasieczka}
\affiliation{Institute for Particle Physics,
             ETH Z\"urich, Switzerland}

\author{Tilman Plehn}
\affiliation{Institut f\"ur Theoretische Physik, 
             Universit\"at Heidelberg, Germany}

\author{Torben Schell}
\affiliation{Institut f\"ur Theoretische Physik, 
             Universit\"at Heidelberg, Germany}

\begin{abstract}
  Top taggers are established analysis tools to reconstruct boosted
  hadronically decaying top quarks for example in searches for heavy
  resonances. We first present a dedicated study of signal efficiency
  versus background rejection, allowing for an improved choice of
  working points. Next, we determine to what degree our mass drop
  selection can be improved by systematically including angular
  correlations between subjets or N-Subjettiness. Finally, we extend the reach
  of the top tagger to transverse momenta below the top mass. This
  momentum range will be crucial in searches for the associated
  production of a Higgs boson with top quarks.
\end{abstract}

\maketitle

\tableofcontents

\newpage

\section{Introduction}
\label{sec:intro}

Because the top quark is the only observed fermion with a weak-scale
mass it can be expected to have strong ties with the mechanism of
electroweak symmetry breaking.  Models which attempt to solve the
hierarchy problem, like supersymmetry, top color, or little Higgs
models~\cite{review}, predict additional states in the top sector to
ameliorate the effect of the top quark on the Higgs boson's mass. In
the Higgs sector we can test modifications of the top Yukawa coupling
in the associated production of a Higgs boson with a top
pair~\cite{tth,tth_exp,sfitter}. An example for weakly interacting
physics beyond the Standard Model, which modifies the top Yukawa
coupling and can be used to complete the Standard Model with free
Higgs couplings in the ultraviolet, is a two Higgs doublet
model~\cite{2hdm}. Typical signatures of new physics linked to the top
sector include top partners decaying to top quarks and missing
energy~\cite{meade,heptop,semilep}, heavy resonances decaying to two
boosted tops~\cite{early,tt_resonances}, or single top
production~\cite{single_top}.

The ultimate experimental goal at the LHC is to search for such
effects in different decay channels, including the purely hadronic
decay of two top quarks. In purely hadronic analyses we first have to
overcome the multi-jet QCD background~\cite{buckets}. After that
initial step we have to extract the signal out of top pair
backgrounds, using the kinematics of the reconstructed tops. Recently,
we have seen that the most promising phase space region for both of
these two tasks are boosted top decays with $p_{T,t} \gtrsim
m_t$~\cite{tagging_review}. Such a boost will, aside from other
improvements, separate the decay products of two top quarks and offer
an handle against combinatorial
backgrounds~\cite{tth,buckets}.

While hadronic signatures are the target of top taggers, experimental
tests have to be performed with an event sample that can be easily
controlled. Semi-leptonic top pairs can be triggered based on a hard
lepton. The momentum of the lepton will also be strongly correlated
with the transverse momentum of the hadronically decaying top
quark. If necessary, $b$-tags can be used to control $W$+jets
backgrounds. Because the aim of this paper is to improve the top
tagging efficiency in particular towards lower transverse momenta we
will present all of our findings in terms of semi-leptonic top pairs,
so they can be easily reproduced by ATLAS and CMS.\bigskip

The idea of studying the substructure of jets and using this structure
to identify massive hadronically decaying particles has been
around for almost 20 years~\cite{seymour,early,bdrs}.  In the
following we will focus on tagging top
quarks~\cite{tth,hopkins,template,pruning,trimming,heptop,scet,recent}
with the help of the \textsc{HEPTopTagger} and in a moderately boosted
regime with $p_{T,t} \lesssim 800$~GeV.  In this framework ATLAS has
published promising results~\cite{atlas}, motivating this detailed
study of top tagging in the upcoming 13~TeV run. To identify and
reconstruct hadronic top decays with higher boost there exist specific
\textsc{HEPTopTagger} improvements~\cite{heptop2}. Our study will be
based on semi-leptonic top pair events which allows for relatively easy
experimental tests in data.\bigskip

As a first step we will introduce some minor modifications of the
\textsc{HEPTopTagger} algorithm in Section~\ref{sec:default}. We will
mostly target a possible shaping of QCD backgrounds towards an
unphysical peak in the reconstructed top mass at low and high jet
multiplicities. For this new default setup we will for the first time
introduce receiver operating characteristics (ROC) curves to quantify
the performance of the tagger beyond the known working points.  In
Section~\ref{sec:angular} we will extend the available observables in
the tagging procedure by angular correlations, given in terms of
Fox--Wolfram moments~\cite{fwm,fwm_higgs}. In Section~\ref{sec:subjettiness}
we will, in a similar spirit, test the combination of the \textsc{HEPTopTagger}
with $N$-Subjettiness. Finally, in
Section~\ref{sec:low_pt} we will modify the tagging algorithm such
that we can identify and reconstruct boosted tops down to $p_{T,t} =
150$~GeV, as suggested by the \textsc{MadMax} study of the $t\bar{t}H$
process~\cite{madmax}.

\section{The new default}
\label{sec:default}

In this first section we test two modifications of the
\textsc{HEPTopTagger} algorithm~\cite{heptop}. None of them will
significantly change the performance of the tagger in terms of signal
efficiency and background rejection. However, they are relevant for
the reconstructed $m_t^\text{rec}$ distribution for the QCD multi--jet
background. Our signal event sample are semi-leptonic decays ($\ell =
e,\mu$) of $t\bar t$ pairs with up to two additional matrix element
jets, generated with \textsc{Alpgen}~\cite{alpgen} and
\textsc{Pythia}~\cite{pythia} using \textsc{Mlm} merging~\cite{mlm}.
For the background we consider a leptonically decaying $W$ plus 2 to 4
matrix element jets simulated the same way.  On the generator level
all hard jets have to to fulfill $p_{T,j} > 25$~GeV, $|\eta_j| < 5$,
and $\Delta R_{jj} > 0.4$. The top mass used for simulation is
173~GeV.  We initially study the current collider energy of $\sqrt{s}
= 8$~TeV and will then move to $\sqrt{s} = 13$~TeV.

Throughout our analysis the top tagging algorithm operates on fat jets
with the Cambridge--Aachen size $R_\text{fat} = 1.8$, reconstructed by
\textsc{Fastjet}~\cite{ca_algo,fastjet}. In this section we require
fat jets with $p_{T,\text{fat}} > 200$~GeV and $|\eta_\text{fat}| <
2.5$. The increased size as compared to the old default of
$R_\text{fat} = 1.5$ will allow us to increase the tagging efficiency
for moderate boost. This increase is triggered by the encouraging
experimental studies of pile-up in filtered fat jets~\cite{pileup},
but it might lead to a re-adjustment of the filtering parameters.  For
consistency reasons we only accept tagged tops with $p_{T,\text{tag}}
> 200$~GeV in this first step. We give all other tagging parameters in
Table~\ref{tab:tagger_setup}: the mass drop required for a massive
splitting is defined by $\text{min} \; m_{j_{1,2}}/m_j <
f_\text{drop}= 0.8$. The jet algorithm stops at subjets with
$m_\text{min} = 30$~GeV, where this parameter can be easily adapted to
a more challenging detector environment.  The $W$ and top masses are
defined on a filtered set of subjets allowing for $N_\text{filt} = 5$
objects, \textit{i.e.}  including up to two jets from final state
radiation.  The kinematic conditions are parameterized by the
\textsc{HEPTopTagger} variables $R_\text{min,max} = (0.85-1.15) \times
m_W/m_t$, $\arctan m_{13}/m_{12} = 0.2-1.3$, and $m_{23}/m_{123} >
0.35$. The selected region has an $A$-like shape in the two
dimensional plane~\cite{heptop}.  Finally, the mass window of the
reconstructed top mass is $150 - 200$~GeV, while in the old default
setup we do not apply a specific cut on the ratio of the reconstructed
$W$ and top masses.\bigskip

\begin{table}[t]
  \centering
  \begin{tabular}{l|r||l|r}
    \hline
    $R_\text{fat}$& \qquad 1.8 &
    $R_\text{min,max}$ & $(0.85-1.15) \times m_W/m_t$ \\ \hline
    $f_\text{drop}$ & 0.8 &
    $m_{23}/m_{123}$ and $\arctan m_{13}/m_{12}$ cuts & 0.35, 0.2, 1.3 \\ \hline
    $m_\text{min}$ [GeV] & 30 &
    $m_t^\text{rec}$ [GeV]& 150-200 \\ \hline
    $N_\text{filt}$ & 5 &
    $(m_W/m_t)^\text{rec}/(m_W/m_t)$ & no cut \\ \hline
  \end{tabular}
  \caption{Parameters in the \textsc{HEPTopTagger} algorithm, as
    defined in the text.}
  \label{tab:tagger_setup}
\end{table}

\begin{figure}[t]
  \includegraphics[width=0.38\textwidth]{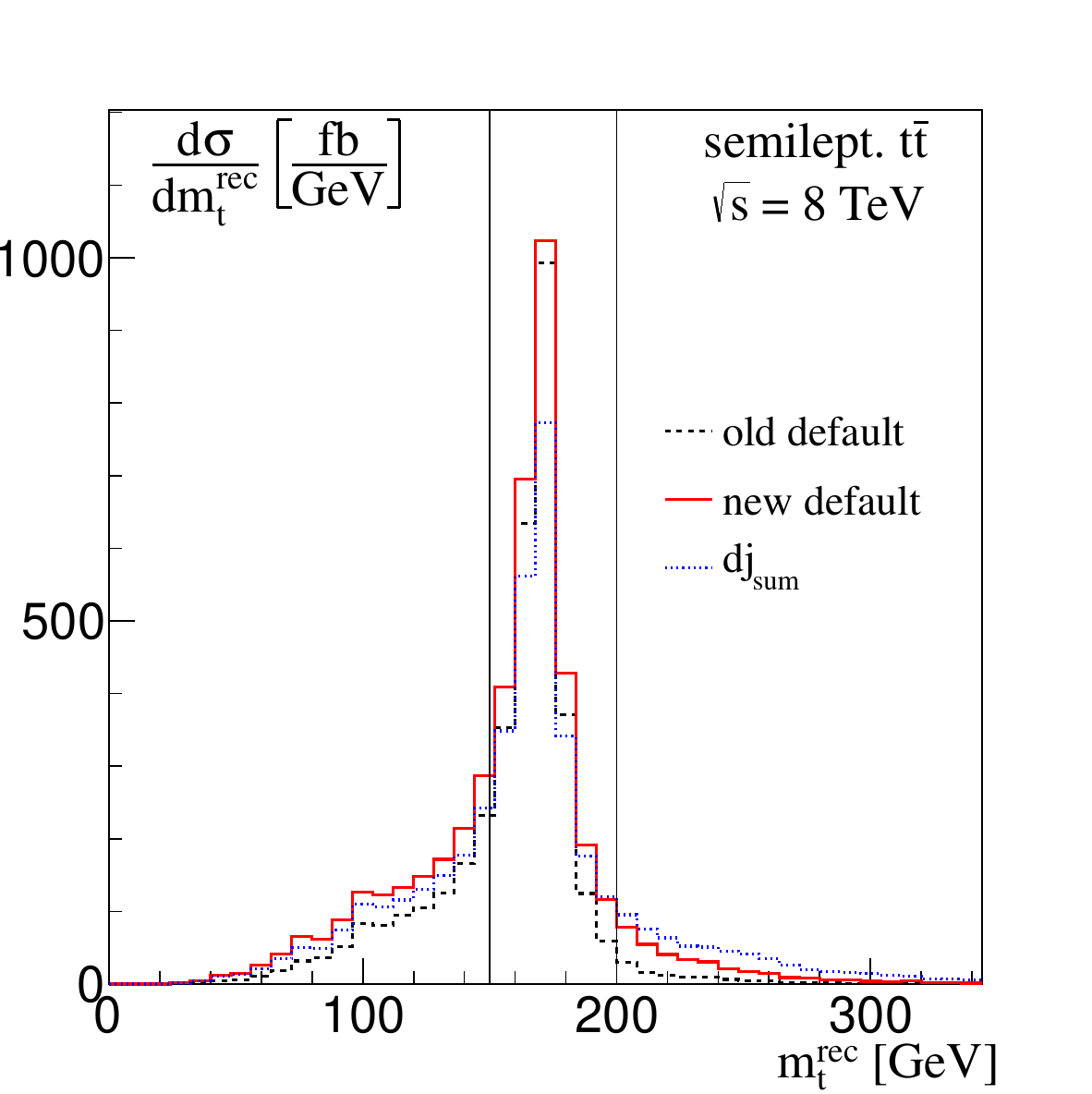}
  \hspace*{0.1\textwidth}
  \includegraphics[width=0.38\textwidth]{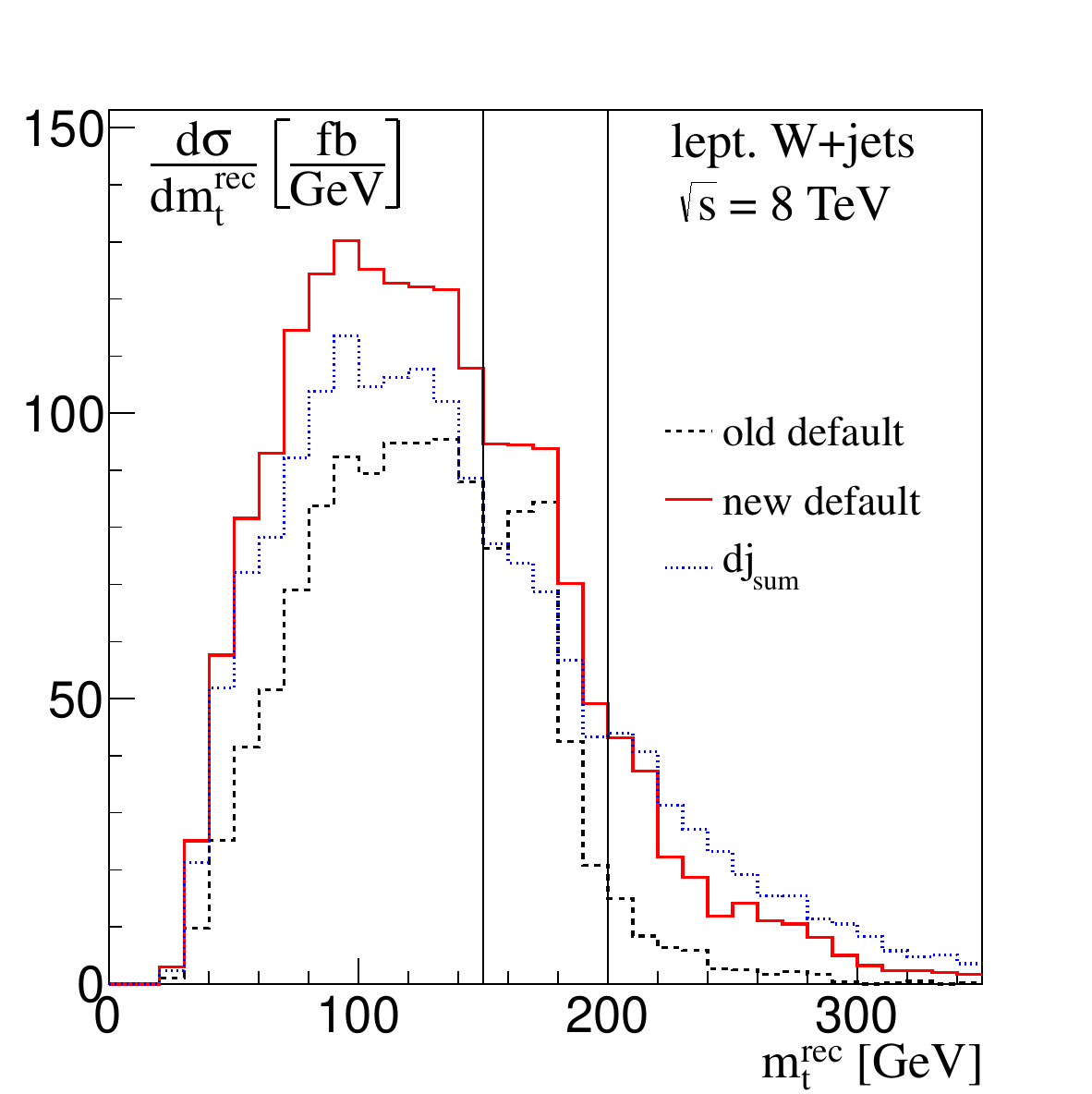}
\caption{Reconstructed top mass for the signal (left) and background
  (right) samples at 8~TeV collider energy. We show the standard order of
  cuts with $|m_{123} - m_t|$ selection (dashed black), the inverted
  order of cuts with $|m_{123} - m_t|$ selection (solid red), and the
  inverted order of cuts with $dj_\text{sum}$ selection (dotted
  blue).}
\label{fig:m_cand}
\end{figure}

Our first modification of the \textsc{HEPTopTagger} algorithm affects
the order in which we apply the top mass selection and the $A$-shaped
$W$ mass constraints. In the standard algorithm we early on select a
filtered triplet of hard subjets closest to the top
mass~\cite{heptop}. This triplet is then required to fulfill the
different $W$ and top mass cuts. The danger is that in the presence of
more than one valid triplet of filtered subjets we pick the wrong one
such that we are guaranteed not to pass the $W$ mass constraints.  To
avoid this, we can first pre-select all triplets which pass the mass
cuts and then pick the one with $m_{123}$ closest to $m_t$ out of
those passing the $W$ mass constraints.  In Fig.~\ref{fig:m_cand}
first see that both versions work on the signal events without an
appreciable difference. The standard ordering slightly shapes the
background around $m_t \sim m_{123}$.\bigskip

The tagging efficiency and mis-tagging rates for both approaches
applied to semi-leptonic top pairs are given in
Table~\ref{tab:default}. The signal efficiencies are defined as the
number of tagged tops divided by the number of hadronically decaying
tops in the event sample with $p_{T,t}>200$~GeV and $|\eta_t| < 2.5$.
For the background we quote the number of mis-tags in the leptonic
$W$+jets sample per number of fat jets passing $p_{T,\text{fat}} >
200$~GeV and $|\eta_\text{fat}|<2.5$. We see that with the new
ordering both the signal efficiency and the background mis-tagging
rate increase, such that typical significances stay constant.  Based
on the reduced shaping of the background mass distributions we will
assume the new order of cuts as \textsc{HEPTopTagger}
standard.\bigskip

\begin{table}[b!]
  \centering
  \begin{tabular}{l|l|r|r|r}
    \hline
    && standard & \multicolumn{2}{c}{inverted} \\ 
    &&  $|m_{123}-m_t|$ & $|m_{123}-m_t|$ & $dj_\text{sum}$ \\ \hline
    \multirow{2}{1cm}{8~TeV}
    &$\eps_S$ & 0.331 & 0.375 & \qquad 0.304 \\ 
    &$\eps_B$ & 0.014 & 0.018 & 0.014 \\ \hline
    \multirow{2}{1cm}{13~TeV}
    &$\eps_S$ & 0.337 & 0.394 & \qquad 0.305 \\ 
    &$\eps_B$ & 0.015 & 0.021 & 0.016 \\ \hline 
    & & old default & new default & \\ \hline
   \end{tabular}
  \caption{(Mis)tagging efficiencies for standard and inverted cut
    order, in the latter case for the $|m_{123} - m_t|$ and
    $dj_\text{sum}$ selections.  All cuts and tagging parameters are
    given in Table~\ref{tab:tagger_setup}.  The new
    \textsc{HEPTopTagger} default setting is indicated.}
  \label{tab:default}
\end{table}

If the algorithm described above is used in a high multiplicity
environment, caused by multi-jet final states and contributions from
pile-up, it is not guaranteed that there is a unique triplet of
filtered subjets which corresponds to the three top decay partons.
The same problem arises for the Higgs tagger in the semileptonic
boosted $t\bar{t}H$ analysis. In that situation choosing the
combination with $|m_{12}-m_H|$ or in this case $|m_{123}-m_t|$ is
guaranteed to shape the background. To construct an alternative
criterion for choosing the correct combination of subjets we start
with the observation that the jet mass can be approximately linked to
the Jade distance~\cite{jade},
\begin{alignat}{5}
m^2_{ij} 
\simeq  E_i E_j \Omega_{ij}^2 
\simeq  p_{T,i} p_{T,j} \; (\Delta R_{ij})^2 \; .
\end{alignat}
We can re-weight the transverse momenta and the geometric separation
to construct an alternative metric for choosing subjet combinations,
like for example the modified Jade distance summed over the three top
decay subjet candidates,
\begin{equation}
dj_\text{sum} = \sum_{(ij)} d_{ij} 
\qqquad \text{with} \quad 
d_{ij} = p_{T,i} p_{T,j} \; (\Delta R_{ij})^4 \; .
\label{eq:mod_jade}
\end{equation}
When we ask for the maximum modified Jade distance we enhance the
weight of the geometric separation as compared to the jet mass,
\textit{i.e.} we prefer those subjet combinations which are more
widely separated. In Fig.~\ref{fig:m_cand} we compare the
$m_t^\text{rec}$ distributions with the old and new
\textsc{HEPTopTagger} default setups.  For the signal as well as the
background sample we find less candidates in the mass window $150 -
200$~GeV. In addition, the background is even less shaped than for the
inverted ordering described above.  The corresponding (mis)tagging
efficiencies are given in Table~\ref{tab:default}. While we find a
tiny improvement in the signal--to--background ratio compared to the
new default setup, the signal efficiency is significantly reduced. We
therefore include this new selection according to
Eq.\eqref{eq:mod_jade} as an option for high--multiplicity
applications of the \textsc{HEPTopTagger}, but do not make it the new
default setting.\bigskip

\begin{figure}[t]
  \includegraphics[width=0.38\textwidth]{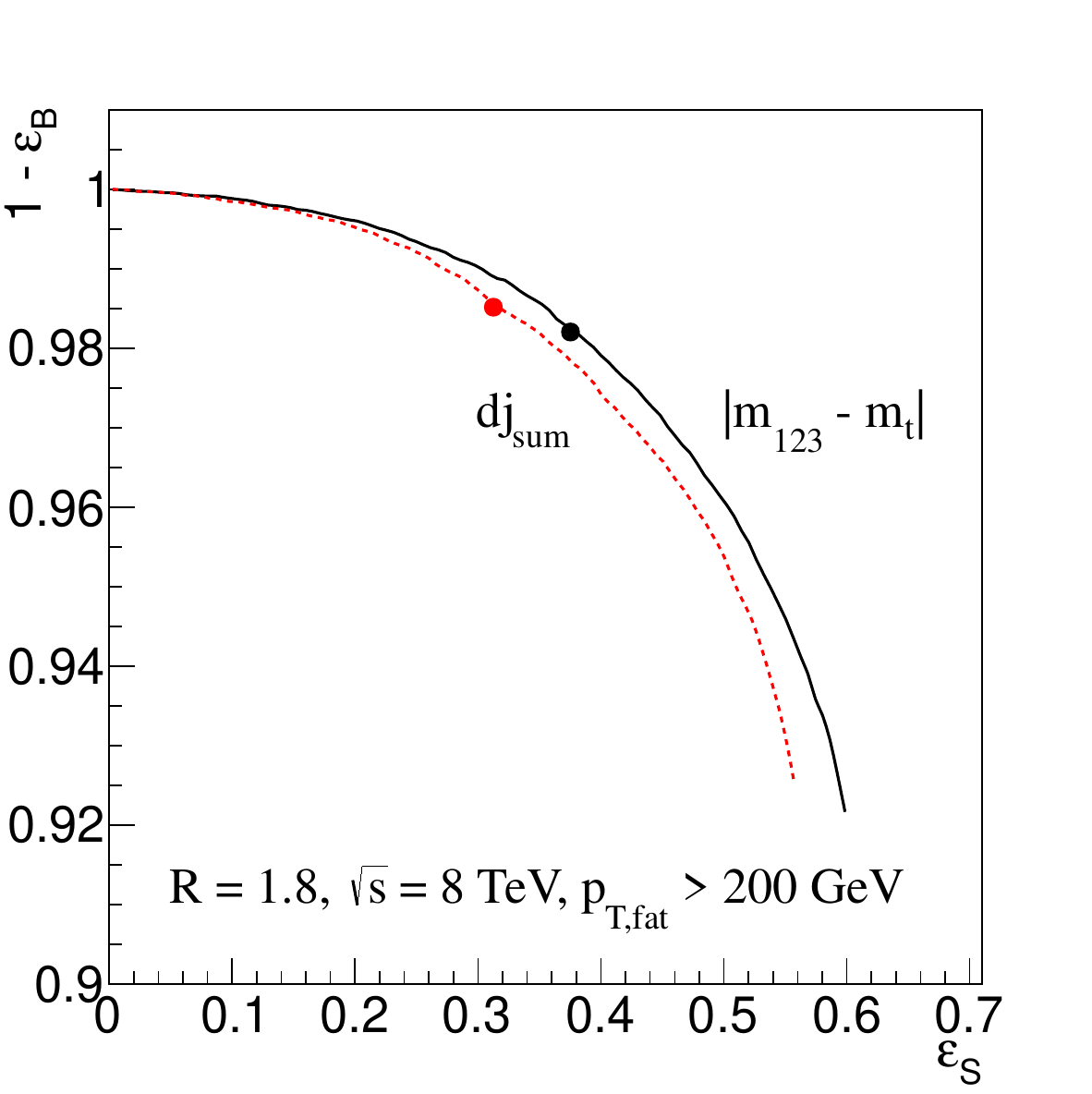}
  \hspace*{0.1\textwidth}
  \includegraphics[width=0.38\textwidth]{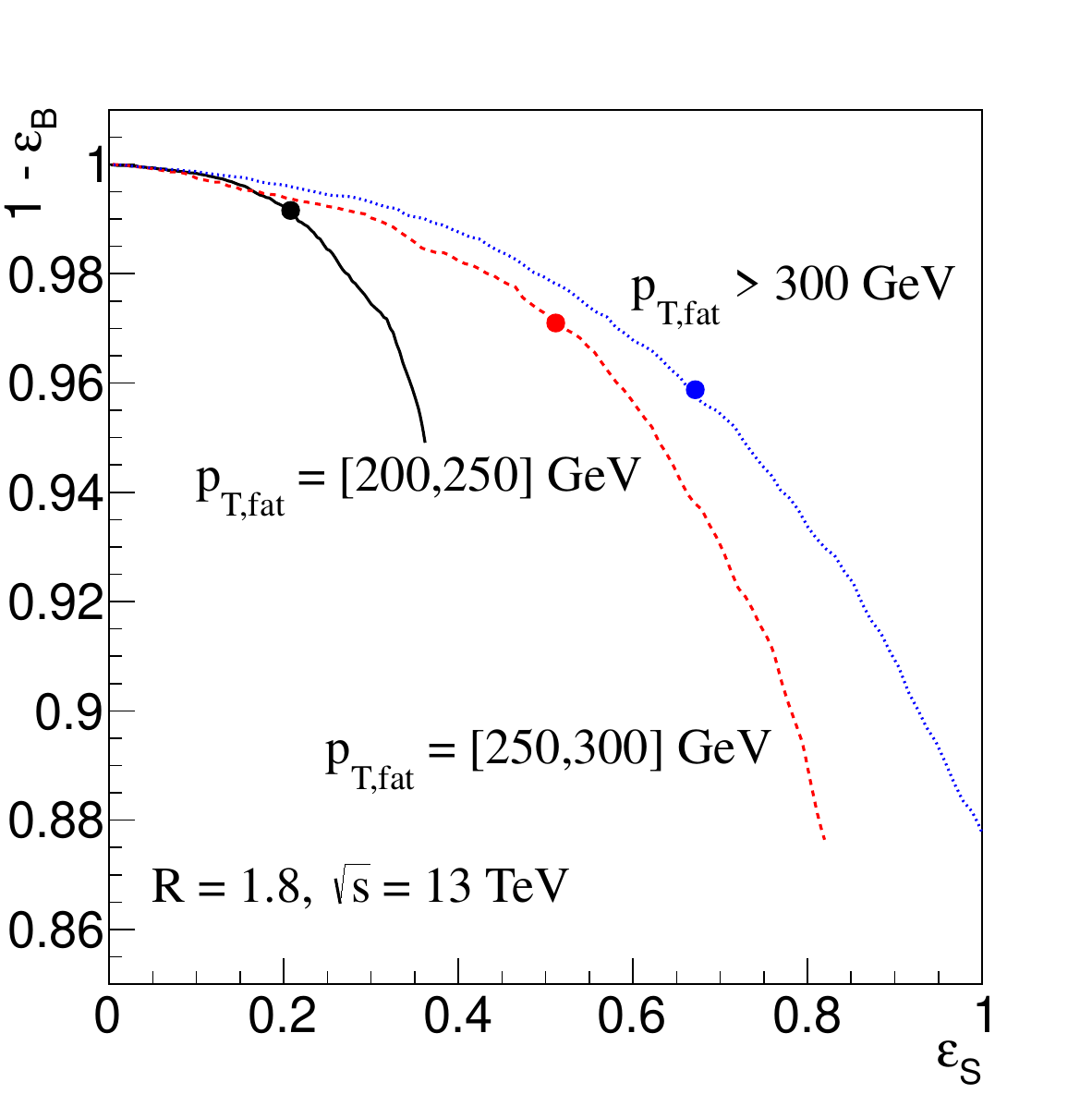}
\caption{ROC curves for the modified \textsc{HEPTopTagger} on
  semileptonic $t\bar{t}$ pairs for 8~TeV (left) and 13~TeV (right)
  collider energy. The standard working points with the cuts given in
  Table~\ref{tab:tagger_setup} are indicated by dots. For 8~TeV we show
  the new default setup and the high-multiplicity modification. For
  13~TeV we quote the performance in slices of $p_{T,\text{fat}}$.}
\label{fig:default_roc}
\end{figure}

To conclusively test the performance of any algorithm, like the
\textsc{HEPTopTagger,} it is not sufficient to only study a single
operating point like the one described in
Table~\ref{tab:tagger_setup}. In particular to test possible
improvements of the tagger~\cite{heptop2} we need to study the
correlation between the signal efficiency $\eps_S$ versus the
background mis-identification probability $\eps_B$. For the standard
working point both of these efficiencies are described in
Table~\ref{tab:default}. The two--dimensional correlation of $\eps_S$
and $1-\eps_B$ are described by a receiver operating characteristics
(ROC) curve. Any point on this curve corresponds to an optimized
parameter setting in the algorithm and can be chosen as the operating
point for an analysis. We derive the ROC curves for the
\textsc{HEPTopTagger} using a boosted decision tree as implemented in
\textsc{Tmva}~\cite{tmva}.  In the \textsc{Tmva} analysis we vary the
the boundaries of the $A$-shaped constraints $\arctan m_{13}/m_{12}$
and $m_{23}/m_{123}$.  To avoid continuously re-running the algorithm
we only allow for tighter cuts than the default algorithm.  The top
mass window which is not part of the actual fat jet algorithm can
change without external constraint, and we include an additional cut
on the reconstructed mass ratio $(m_W/m_t)^\text{rec}$. For more
details on the BDT optimization we refer to Ref.~\cite{fwm_higgs}.

In the left panel of Fig.~\ref{fig:default_roc} we show ROC curves for
the re-ordered \textsc{HEPTopTagger} algorithm for 8~TeV collider
energy. We show results for the new standard setup and the
high-luminosity setup ordered by the modified Jade distance of
Eq.\eqref{eq:mod_jade}.  As mentioned before, the efficiency for the
signal is defined as the number of tagged tops normalized to the
number of hadronically decaying tops with $p_{T,t} > 200$~GeV and
$|\eta_t| < 2.5$. For the background the normalization is given by the
number of fat jets fulfilling the same $p_T$ and $\eta$ criteria.
Computed for the full $t\bar{t}$ sample we reach signal efficiencies
of 40\% for a background rejection over 97.5\%.  When computed over
the entire $p_T$ range of the fat jet or top quark we see that the
algorithm essentially breaks down at signal efficiencies above $\eps_S
= 0.6$. The reason is that the number of possibly tagged tops is
limited by the number of tops with all three decay subjets inside the
fat jet.

For 13~TeV, shown in the right panel of Fig.~\ref{fig:default_roc}, we
break the efficiency into three transverse momentum slices of the top
or the fat jets, namely $p_T = 200-250$~GeV, $p_T = 250-300$~GeV, and
$p_T > 300$~GeV. Higher transverse momenta will be statistically
limited in the $t\bar{t}$ sample and are not the focus of this
study. The signal efficiencies are defined as the number of tags
obtained from the considered fat jets divided by the number of
hadronically decaying tops in the event sample within the given range
of transverse momenta. The combined curve for the $t\bar{t}$ sample
will essentially coincide with the lowest $p_T$ bin, simply because of
the composition of the signal events.  Again, we observe that the
signal efficiency is limited in the soft region.  Above $p_{T,t} =
300$~GeV we can reach almost 100\% signal efficiencies. The
performance of the new default working point from
Table~\ref{tab:tagger_setup} is indicated by a dot. Depending on the
analysis, we see that over the entire $p_T$ range the
\textsc{HEPTopTagger} can reach a QCD background rejection of 99\% for
signal efficiencies between 20\% for soft tops and 40\% for hard
tops. If required, the optimal working point of the top tagger can be
chosen as a function of the fat jet momentum for example to guarantee
a constant signal efficiency or background rejection according to
Fig.~\ref{fig:default_roc}.\bigskip

The one remaining question is if the new version of the
\textsc{HEPTopTagger} still benefits from the pruned reconstructed top
mass~\cite{pruning}, as it does for the old version~\cite{heptop2}. In
terms of the ROC curves shown in Fig.~\ref{fig:default_roc} we can
compare the new default tagger with the version including the pruned
mass. It turns out that the two ROC curves are identical for each
transverse momentum slice. The former improvement based on including
the pruned mass is already taken into account in the improved
efficiencies shown in Table~\ref{tab:default}.

\section{Angular correlations}
\label{sec:angular}

\begin{figure}[b!]
  \includegraphics[width=0.38\textwidth]{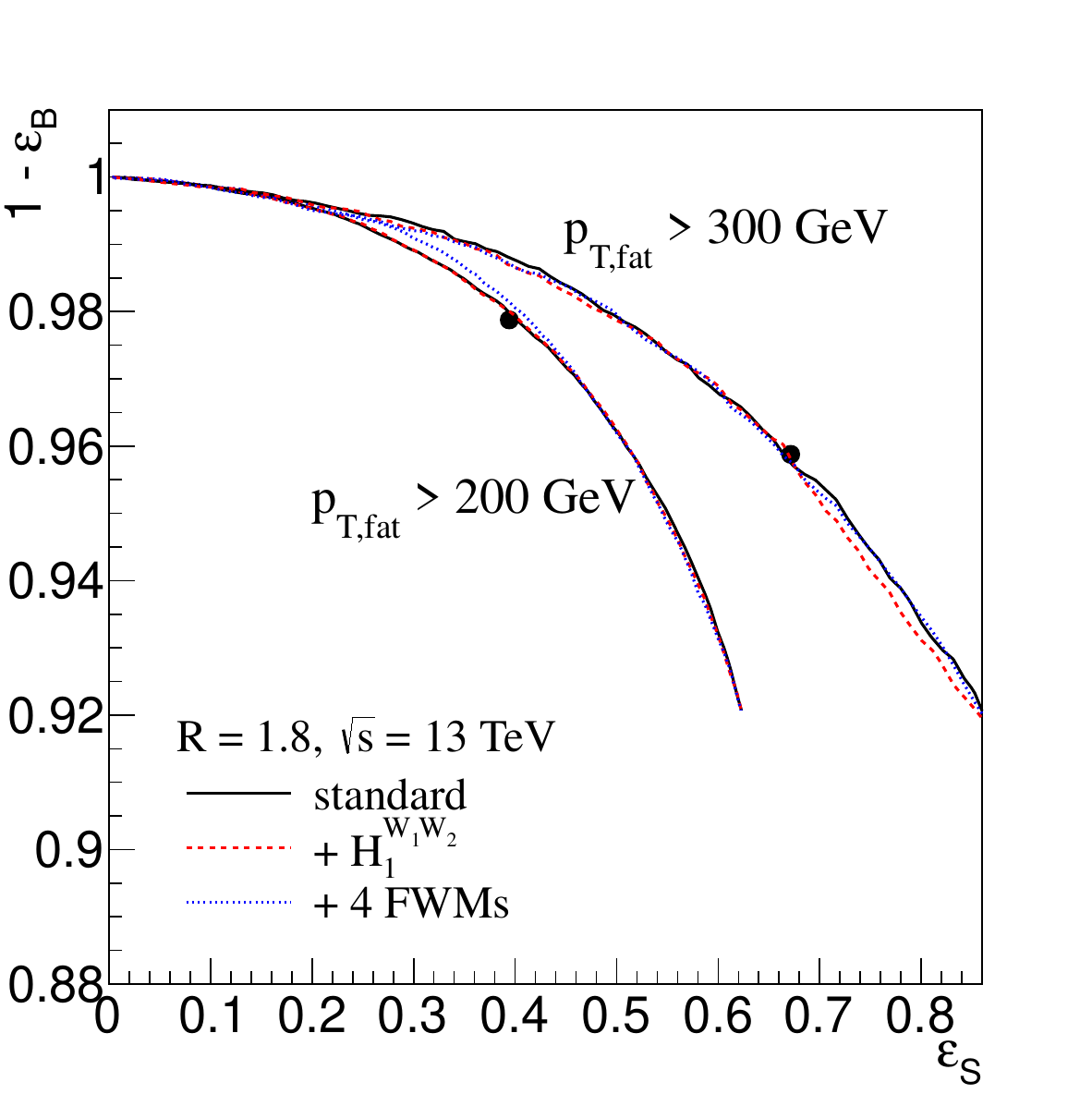}
  \hspace*{0.1\textwidth}
  \includegraphics[width=0.38\textwidth]{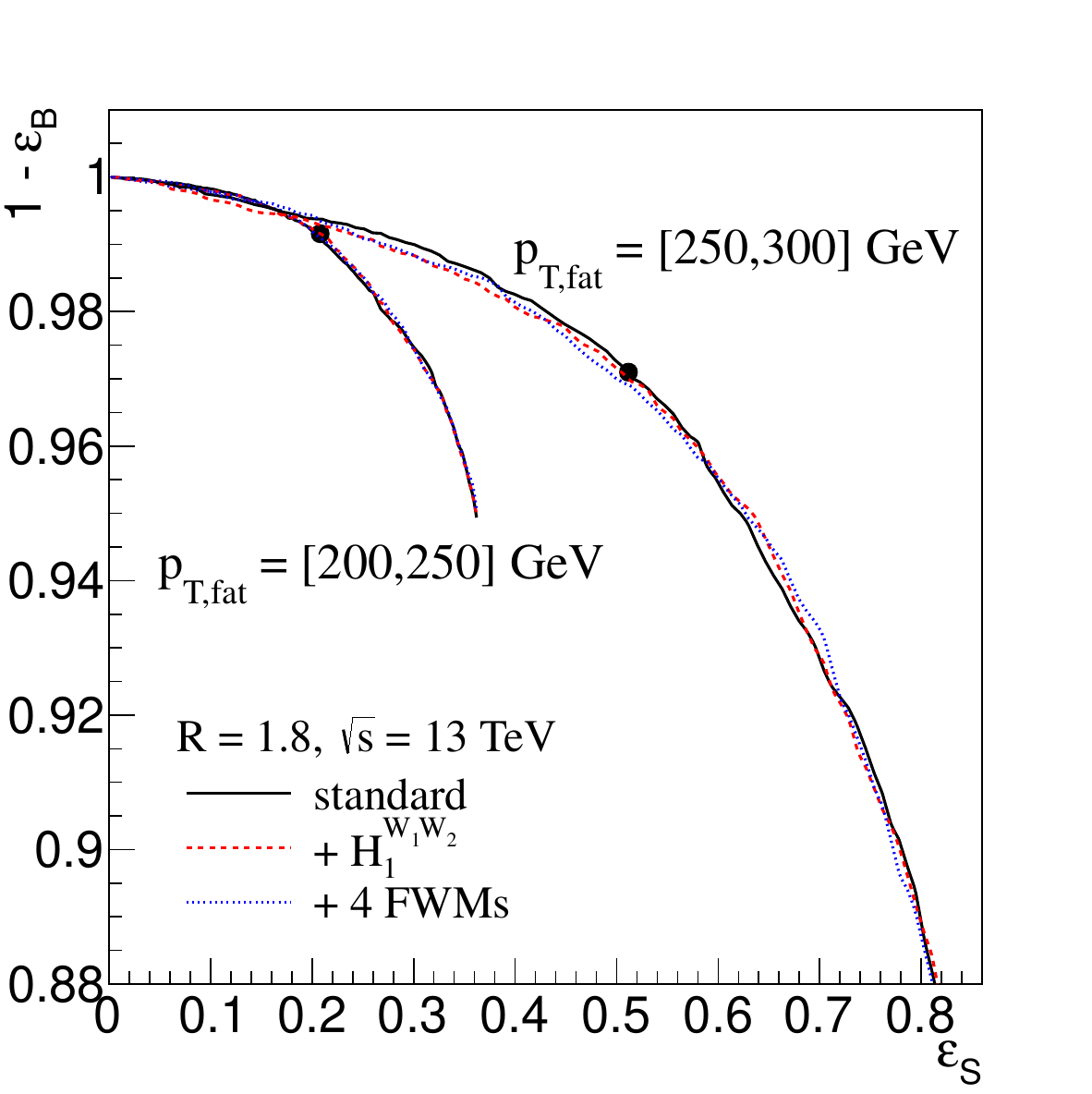}
\caption{ROC curves for the modified \textsc{HEPTopTagger} including
  angular correlations via unit--weight Fox--Wolfram moments in slices
  of $p_{T,\text{fat}}$.  The standard working point from
  Table~\ref{tab:tagger_setup} is indicated by a dot. We assume a
  collider energy of 13~TeV.}
\label{fig:roc_fwm}
\end{figure}

To make a conclusive statement about possible improvements to the
\textsc{HEPTopTagger} by using angular correlations of the subjets we
need a systematic way to include such angular
correlations. Correlations between a pair of subjets or other objects
can be fully described in the basis of spherical harmonics.
Fox--Wolfram moments are then constructed as a sum over all $2\ell+1$
directions, weighted by a free function $W_i$~\cite{fwm}
\begin{alignat}{1}
H_\ell^x = \frac{4\pi}{2\ell+1}
       \sum_{m=-\ell}^\ell \;
       \left| \sum_{i=1}^N  W_i^x \; Y_\ell^m(\Omega_i) 
       \right|^2 \; .
\label{eq:fwm_def1}
\end{alignat}
The index $i$ runs over all $N$ objects. The individual
coordinates $\Omega_i$ can be replaced by the corresponding angular
separation $\Omega_{ij}$, allowing us to write the moments in terms of
Legendre polynomials
\begin{alignat}{2}
H^x_\ell = \sum_{i,j=1}^N \; W_{ij}^x \; P_\ell(\cos \Omega_{ij}) 
\qqquad \text{with} \quad 
W_{ij}^x = W_i^x W_j^x \; .
\label{eq:fwm_def2}
\end{alignat}
Two common weights are the transverse momentum or the unit
weights~\cite{fwm,fwm_higgs}
\begin{alignat}{1}
W_{ij}^T = \frac{p_{T i}\,p_{T j}}{\left(\sum p_{T i}\right)^2}  
\qqqquad 
W_{ij}^U = \frac{1}{N^2} \; .
\label{eq:fwm_weight}
\end{alignat}
The advantage of the transverse--momentum weight is that it suppresses
soft and collinear jets, which are hard to correctly describe in
QCD. However, tests show us that inside a top tagger the unit weight
is more promising. Therefore, we do not use the transverse momentum
weight in this analysis. We could in principle also compute these
moments over other objects inside the fat jet, for example before
filtering or the five subjet structures before re-clustering. However,
all of this would make our results process dependent and hence not
suitable for this study.

Using the unit weight the Fox--Wolfram moments analyze the angular
correlations inside the fat jet. We evaluate them in the top rest
frame, to remove the leading effect from the boost. This means that
for the unit weights we do not have to define a reference axis and
only two of the three angles between the subjets are independent. The
fourth direction is given by the direction of the fat jet or the boost
into the top rest frame.\bigskip

To quantify the improvement which can be gained from the angular
correlation we evaluate the Fox--Wolfram moments for each combination
of subjets using a boosted decision tree as implemented in
\textsc{Tmva}. For the angular analysis we label the two subjets which
reconstruct $m_W$ best as $W_1$ and $W_2$, ordered by transverse
momentum. The remaining subjet is then labeled the $b$-jet. As long as
we include a full set of correlations in the Fox--Wolfram moments,
these names do not matter for the performance. They will only become
relevant once we interpret the results. Using \textsc{Tmva} we
determine the most decisive moments for each combination of two to
four subjets and the boost direction, \textit{i.e.} two spatial
directions out of $\{ W_1, W_2, b, \vec{p}_\text{boost} \}$. This
allows us to define a limited set of Fox--Wolfram moments which we can
reliably compare to the purely kinematics--based selection discussed
before.

In Fig.~\ref{fig:roc_fwm} we show ROC curves for the full event sample
and for slices in the transverse momentum of the fat jet. The purely
QCD-inspired selection criteria of Table~\ref{tab:tagger_setup} is
contrasted with the selection including the available angular
correlations. The moment with the strongest separation for signal and
background is the first moment $H^U_1$ built from the two $W$ decay
jets. However, for all four $p_{T,\text{fat}}$ choices we see that
including this additional information has no effect on the performance
of the tagger.

In addition, we show results including the leading moments from the
best sets of momenta. As mentioned above, the leading moment is $H^U_1$
from the $(W_1,W_2)$ selection. Other examples for highly ranked
moments are $H^U_2$ from the $(W_1,W_2,\vec{p}_\text{boost})$ selection
or $H^U_1$ defined over $(W_2,b,\vec{p}_\text{boost})$. The results from
an optimized tagger using the best four moments are also shown in
Fig.~\ref{fig:roc_fwm}, again with no visible improvement.\bigskip

The bottom line of this systematic analysis of angular correlations in
top tagger is that the angular correlations among the top decay
products and the top direction are already included in the QCD--based
selection. Apparent improvements are within the reach of the ROC
curve. For the given transverse momentum range adding more angular
information does not lead to a measurable improvement of the
\textsc{HEPTopTagger}.

\section{N-Subjettiness}
\label{sec:subjettiness}

\begin{figure}[b!] 
  \includegraphics[width=0.38\textwidth]{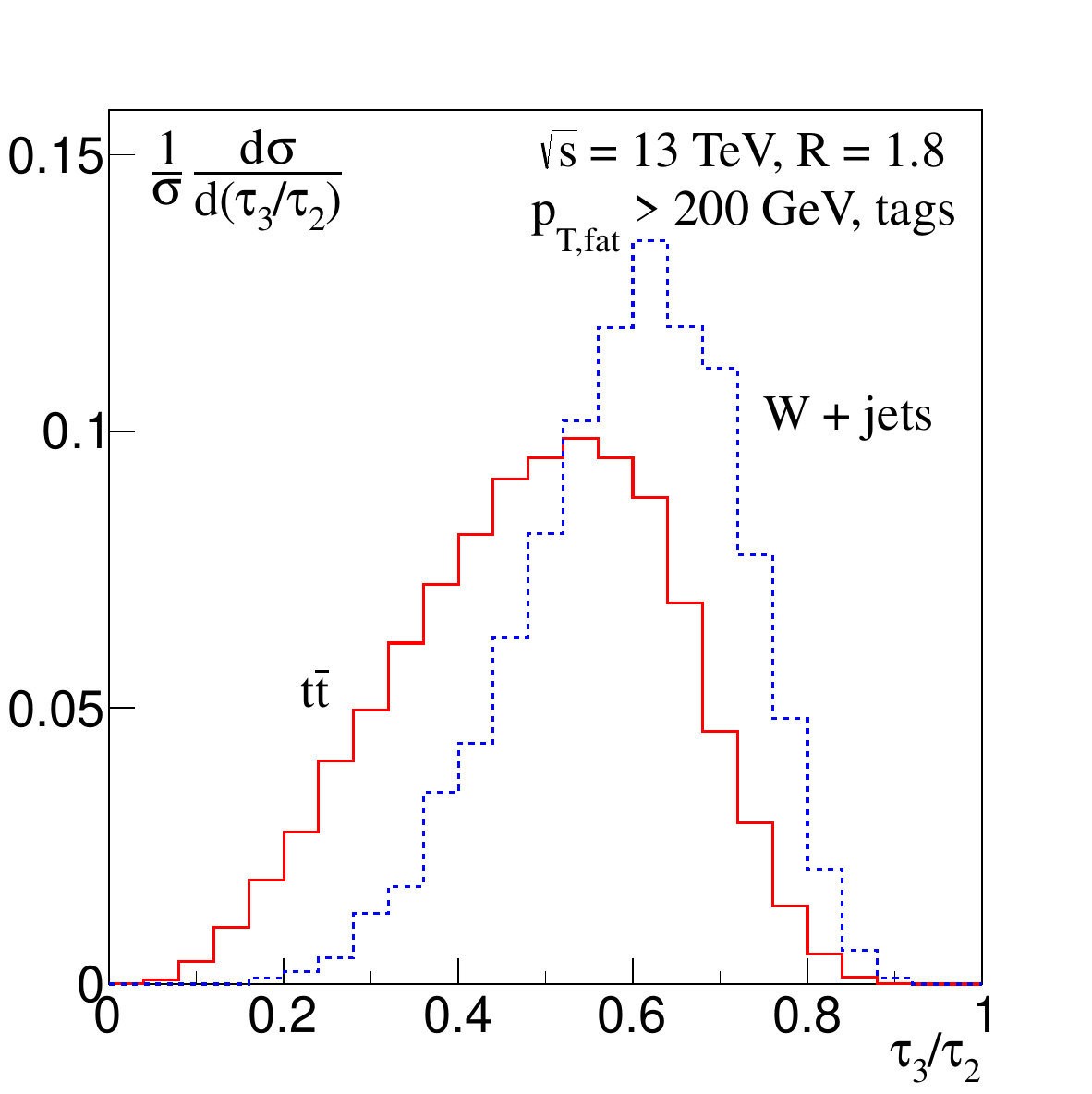}
  \hspace*{0.1\textwidth}
  \includegraphics[width=0.38\textwidth]{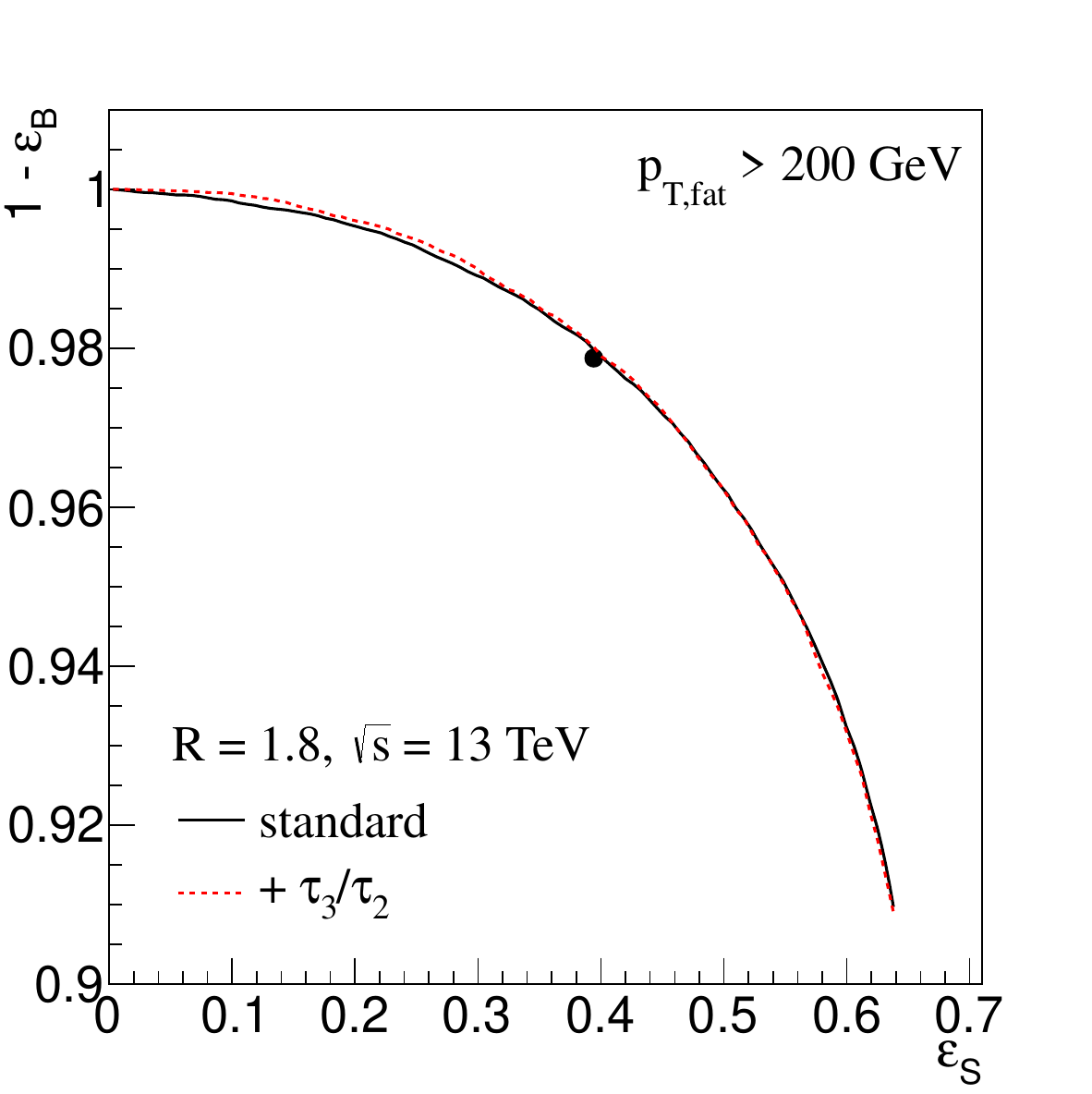}
  \caption{Left: $\tau_3/\tau_2$ for tagged fat jets The red solid
    curves show the $t\bar{t}$ signal while the blue dotted curves
    give the $W$+jets background. Right: ROC curves for the combined
    \textsc{HEPTopTagger} and $\tau_3/\tau_2$.}
  \label{fig:subjettiness}
\end{figure}

A second way to identify boosted hadronically decayed top quarks could
be a combination of a mass--drop criterion with
$N$-Subjettiness~\cite{nsubjettiness}.  This additional observable
measures how well a fat jet is described by a given number of
subjets. It starts by constructing $N$ reference axes and then
measures how well the $k$ fat jet constituents fit to those axes,
\begin{equation}
 \tau_N = \frac{1}{R_0 \sum_k p_{T,k}} \sum_k p_{T,k} \min \left(
   \Delta R _{1,k}, \Delta R _{2,k}, \cdots, \Delta R _{N,k} \right) \;.
 \label{eq:tau_N}
\end{equation}
$\Delta R$ is the usual geometric separation and $R_0$ is an intrinsic
cone size, usually the value used to construct the fat jet. Small
values of $\tau_N$ indicate that the jet is consistent with fewer than
$N$ subjets. Consequently, ratios of the kind $\tau_N/\tau_{N-1}$
allow us to distinguish between jets described by $N$ or $N-1$
substructures. For top tagging the ratio $\tau_3/\tau_2$ is
expected to be most useful.

Similar to the case of angular correlations studied in the previous
section, we combine the \textsc{HEPTopTagger}~\cite{heptop} setup
given in Table~\ref{tab:tagger_setup} with $N$-Subjettiness based on
axes using the one-pass-$k_T$ method implemented in
\textsc{SpartyJet}~\cite{spartyjet}.\bigskip

In the left panel of Fig.~\ref{fig:subjettiness} we show the $\tau_3 /
\tau_2$ distributions for the fat jets tagged by the new default
\textsc{HEPTopTagger}. We see that $\tau_3 / \tau_2$ without any
additional cuts carries information that can be used to distinguish
between signal and background.  To check the power of this additional
variable in addition to the tagging algorithm we again compute the
corresponding ROC curve with a variable combination of tagging
parameters and $N$-Subjettiness using boosted decision trees as
implemented in \textsc{Tmva}~\cite{tmva}.  We show the results in the
right panel of Fig.~\ref{fig:subjettiness}: first, we show the new
default \textsc{HEPTopTagger} described in Section~\ref{sec:default},
allowing for an optimized mass window $m_{123}$ as well as tighter
$A$-shaped mass plane cuts and a cut on $m_W/m_t$. We then include
$\tau_3/\tau_2$ in the boosted decision tree and see that it gives a
hardly visible improvement in the low efficiency range.\bigskip

The bottom line of this analysis of a combined mass--drop and
$N$-Subjettiness tagger is again that the additional information
includes relevant information, but that it does not lead to an
improvement compared to an optimized \textsc{HEPTopTagger} setup.  The
apparent improvement corresponds merely to a shift in the working
point of the mass drop tagger.

\section{Low boost}
\label{sec:low_pt}

The lesson we learned in Sections~\ref{sec:angular} and~\ref{sec:subjettiness}, namely
that adding angular correlations or additional observables to the tagging algorithm has
essentially no effect, is limited to sufficiently boosted top quarks
for which the algorithm has the chance to identify all three decay
jets from mass drop criteria. In the range of $p_{T,t} = 150 - 200$~GeV
this might not be the case, while the fat jet still includes most of
the kinematic information from the top decay.  For example, the
softest decay jet might be pushed outside the fat jet and replaced by
an initial state QCD jet, but the two leading jets could still be used
to identify a massive top decay. For such events the mass drop
criterion will not be efficient enough to separate the signal from the
background, which brings us back to angular correlations between the
decay products. 

\begin{table}[b!]
  \centering
  \begin{tabular}{l|r|r}
\hline
$\sigma_\text{tot} = 360$~fb & \qquad all $p_{T,H}$ & $p_{T,H} > 100$~GeV \\ \hline
all $p_{T,t}$                & 100\%         &  48\% \\
$p_{T,t_1} > 150$~GeV        &  59\%         &  37\% \\
$p_{T,t_1} > 200$~GeV        &  39\%         &  27\% \\  
$p_{T,t} > 150$~GeV          &  29\%        &   16\% \\
$p_{T,t} > 200$~GeV          &  15\%        &  8.8\% \\ \hline
  \end{tabular}
  \caption{Fraction of the total cross section for $t\bar{t}H$
    production at 13~TeV, passing a range of cuts on the transverse momentum
    of the harder top ($p_{T,t_1}$) or both tops ($p_{T,t}$).}
  \label{tab:tth}
\end{table}

\begin{table}[t]
\centering
\begin{tabular}{l|r|r|r|r}
\hline
&\multicolumn{2}{c|}{default}&\multicolumn{2}{c}{low-$p_T$ mode}\\
 & (mis)tags [fb] & fraction & (mis)tags [fb] & fraction \\ \hline
type-1   & 5309 & 57\% & 5967 & 52\% \\
type-2   & 1283 & 14\% & 1863 & 16\% \\
type-3   & 2712 & 29\% & 3601 & 32\%\\ 
$\eps_S$ & 0.287& &0.353& \\ \hline
$W$+jets  & 1200 & & 1663 & \\ 
$\eps_B$ & 0.007& &0.010& \\ \hline 
\end{tabular}
\caption{Comparison of tagging results in the new default setup without (left)
  and with the low-$p_T$ mode (right). All tags fulfill
  $p_{T,\text{fat}} >150$~GeV as well as
  $p_{T,\text{tag}}=150-200$~GeV. The working point for the low-$p_T$ mode
  is given in Table~\ref{tab:lowpt_cuts}. The quoted cross section
  values correspond to the semi-leptonic $t\bar{t}$ sample.}
\label{tab:lowpt_compare}
\end{table}

One obvious application for a top tagger at low momenta is the
associated production of a Higgs with a pair of top quarks~\cite{tth}.
In Table~\ref{tab:tth} we show the fraction of events for this process
including one or two top quarks with $p_{T,t} > 150$~GeV or $p_{T,t} >
200$~GeV. Reducing the transverse momentum threshold by 50~GeV
increases the number of events by 50\% for one top tag and by 100\%
for two top tags.\bigskip

The default \textsc{HEPTopTagger} has never targeted top quarks with
$p_{T,t} < 200$~GeV. This consistency condition is related to the size
of the fat jet and based on the assumption that the algorithm has to
be able to see all three top decay subjets~\cite{heptop}. In this
section we suggest a new analysis step targeting events with
$p_{T,t}^\text{rec} = 150 - 200$~GeV. In this range we will employ
angular correlations through the Fox--Wolfram moments, introduced in
the last section.

As before, our study is based on semi-leptonically decaying top
pairs. The size of the fat Cambridge--Aachen jet is $R_\text{fat} =
1.8$, and it is required to have $p_{T,\text{fat}} > 150$~GeV and
$|\eta_\text{fat}| < 2.5$. Those fat jets are analyzed with the (new)
default \textsc{HEPTopTagger} described in Section~\ref{sec:default}.
To determine the quality of each top tag with $p_{T,t}^\text{rec} =
150 - 200$~GeV we rely on the geometric separation of the
parton--level top decay jets and the reconstructed top constituents
for a given mapping $j_i$~\cite{heptop2},
\begin{alignat}{5}
\Delta R_\text{sum}^2 = \sum_{i=1}^3 \Delta
R^2(p_i^\text{rec},p_{j_i}^\text{parton}) \; .
\end{alignat}
We minimize $\Delta R_\text{sum}^2$ to define the best mapping of
parton level and reconstructed top decay products. Based on the parton
level information we can then assign it to one of three types:
\begin{itemize}
\item[] type-1: all three subjets of the tagged top quark correspond to
  top decay products at parton level,
\item[] type-2: the two hardest subjets correspond to top decay
  products, the third subjet does not,
\item[] type-3: everything else.
\end{itemize}
This way the type-3 category includes events where one or zero subjets
correspond to top decay products, but also events where only the
hardest subjet cannot be assigned to a top decay parton.  In
Table~\ref{tab:lowpt_compare} we see the fraction of type-1 to type-3
tags after requiring a reduced threshold $p_{T,\text{fat}} > 150$~GeV
and only considering a low-$p_T$ slice with $p_{T,t}^\text{rec} = 150
- 200$~GeV. For the default tagger with these two modifications 57\%
of the additional tags are of type-1, \textit{i.e.}  all top decay
products can be assigned to parton level information. For an
additional 14\% of events the two leading subjets can be linked to top
decay product at parton level, while for 29\% this matching is
problematic.\bigskip

\begin{table}[b!]
\centering
\begin{tabular}{l|r|r}
\hline  
& min & max \\ \hline
$m_t^\text{rec}$ [GeV] & 108. & 282. \\ 
$(m_W/m_t)^\text{rec}/(m_W/m_t)$ & 0.717 & 1.556 \\ \hline
$\arctan\left(m_{13}/m_{12}\right)$ & 0.441 & 0.889 \\ 
$m_{23}/m_{123}$ & 0.412 & 0.758 \\ \hline
$H^{W_1W_2}_1$ & 0.048 & 0.373 \\ 
$H^{pW_1W_2}_2$ & 0.019 & 0.524 \\ 
$H^{pbW_1W_2}_2$ & 0.044 & 0.276 \\ 
$H^{pbW_2}_1$ & 0.145 & 0.445 \\ \hline
\end{tabular}
\caption{Cuts defining the working point used for the low-$p_T$ mode
  shown in Fig.~\ref{tab:lowpt_compare}. The values are extracted
  using simulated annealing.}
\label{tab:lowpt_cuts}
\end{table}

The critical step at which not perfectly matched top decays in the
type-2 and type-3 categories fail the default tagging algorithm is the
final mass window $m_t^\text{rec} = 150-200$~GeV. In the modified
low-$p_T$ mode the \textsc{HEPTopTagger} first accepts all tagged
tops with $m_t^\text{rec} = 150-200$~GeV, as long as they fulfill the
now lower consistency criterion $p_{T,\text{tag}} = 150-200$~GeV.  

\begin{figure}[t]
  \includegraphics[width=0.38\textwidth]{./figs/roc_bdt_13tev_pt}
  \hspace*{0.1\textwidth}
  \includegraphics[width=0.38\textwidth]{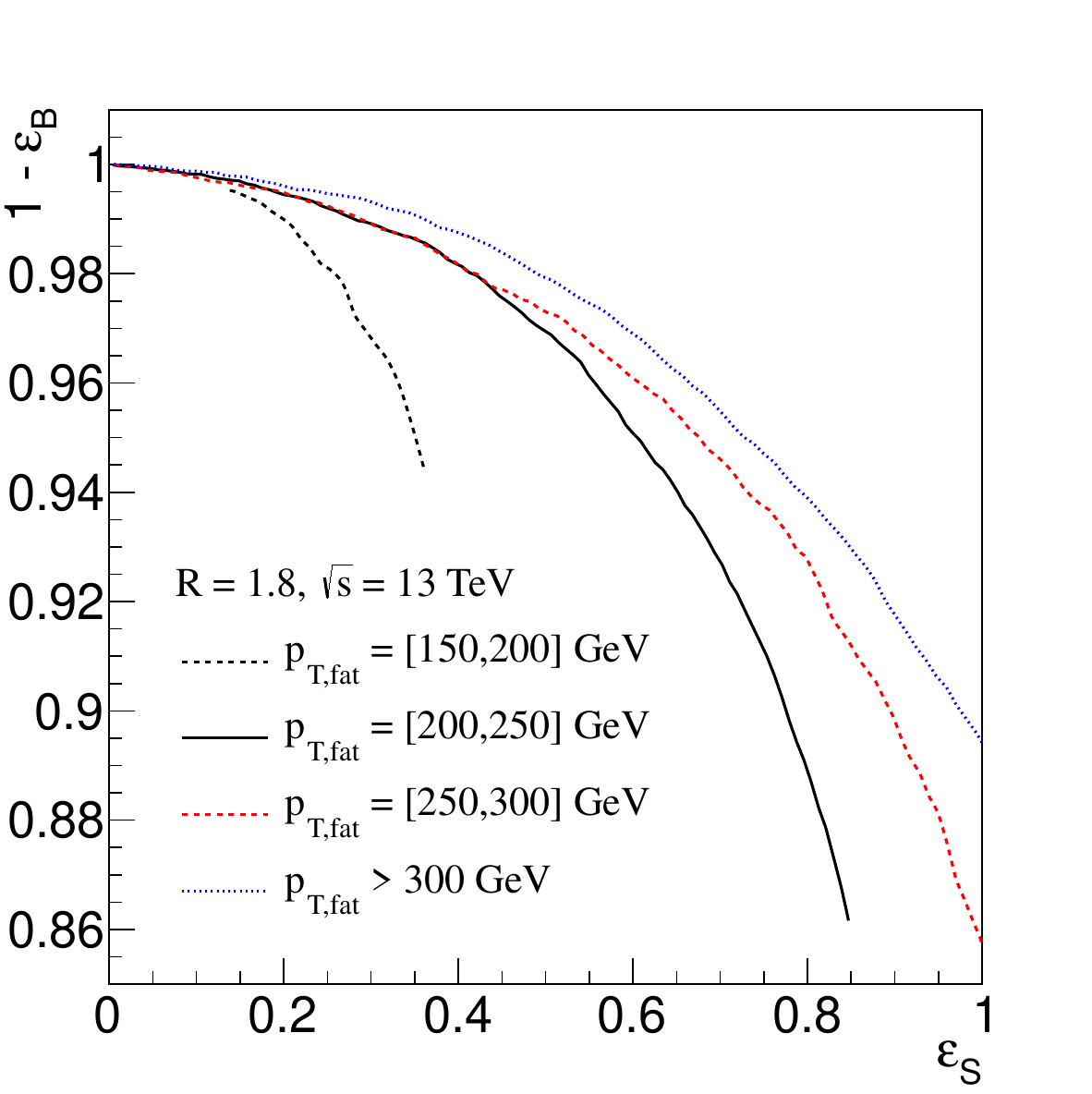}
\caption{ROC curves for the new standard \textsc{HEPTopTagger} without
  (left) and with the new low-$p_T$ mode (right) in slices of
  $p_{T,\text{fat}}$. The signal performance is given for semileptonic
  $t\bar{t}$ pairs at 13~TeV collider energy.}
\label{fig:lowpt_roc}
\end{figure}

\begin{figure}[b!]
 \includegraphics[width=0.38\textwidth]{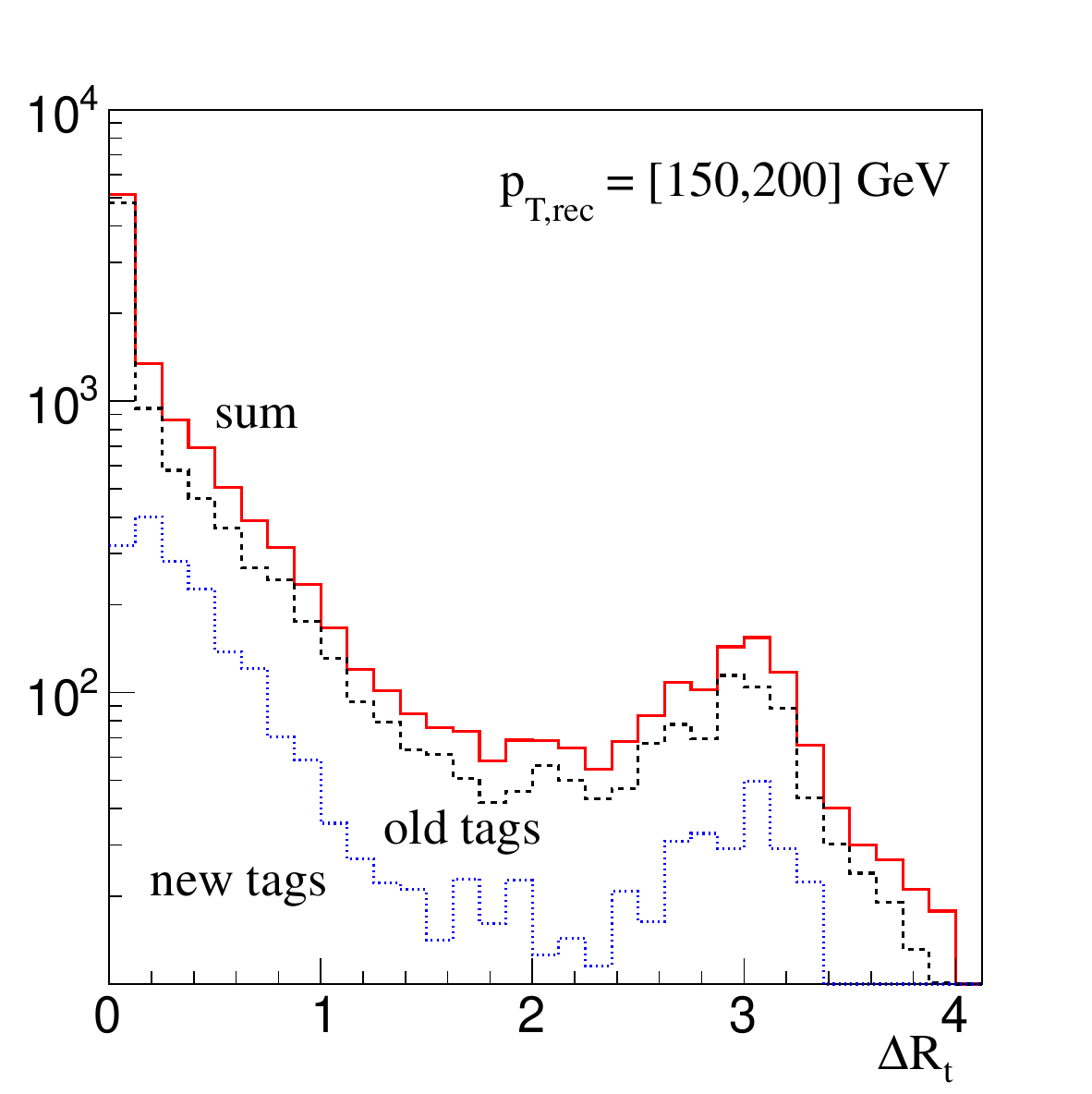}
  \hspace*{0.1\textwidth}
 \includegraphics[width=0.38\textwidth]{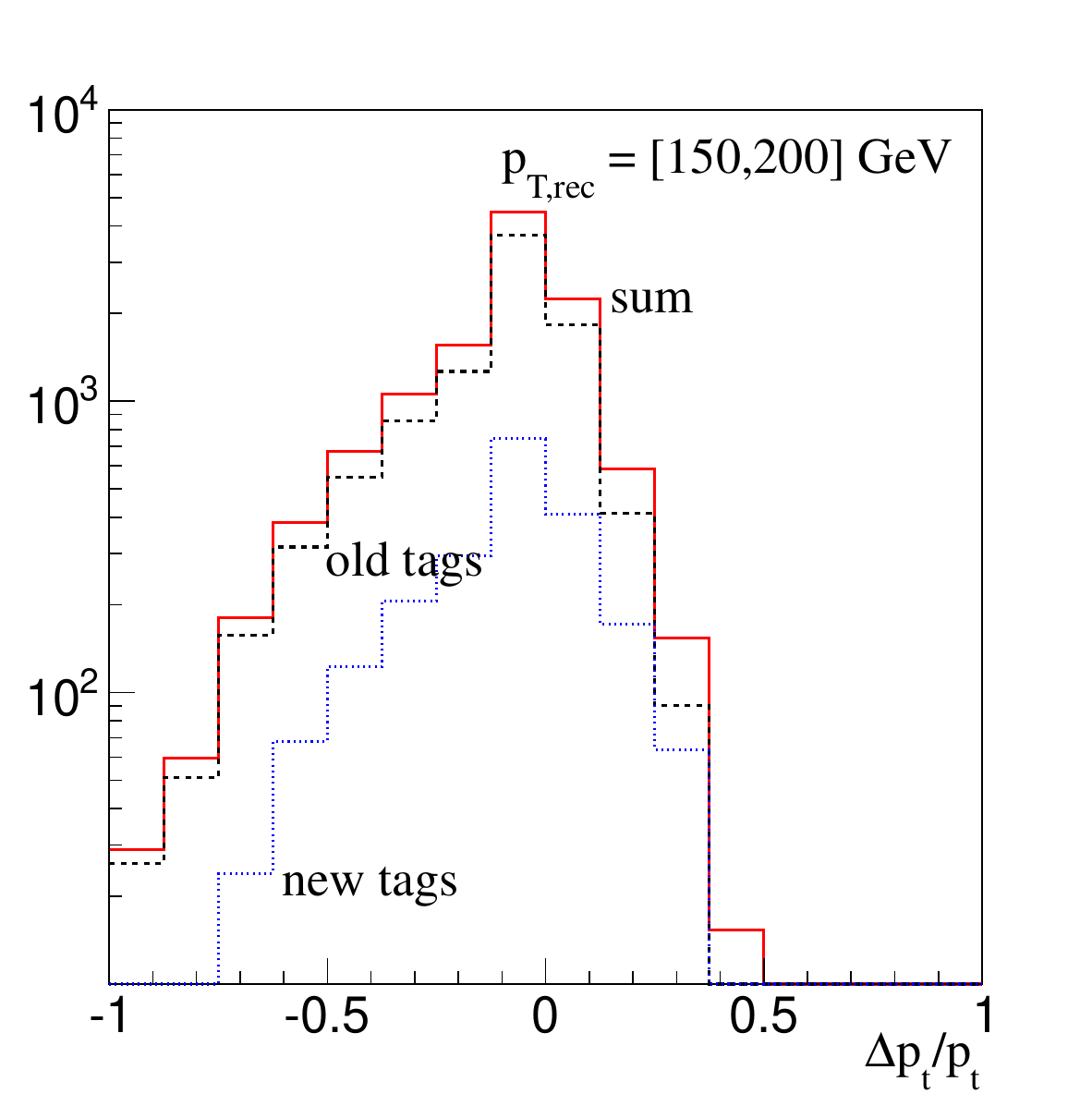}
 \caption{Reconstruction for the original tags (black), the additional
   tags in the low-$p_T$ mode (blue), and all events in the
   semi-leptonic $t\bar{t}$ sample (red). We show the angular
   difference and the relative difference in transverse momentum
   between the reconstructed top momentum and parton level truth for
   all tagged tops.}
\label{fig:reco}
\end{figure}

We then target all
events which do not pass the $m_t^\text{rec}$ condition but 
fall within the transverse momentum range $p_{T,\text{tag}} = 150-200$~GeV.  
For them we widen the $m_t^\text{rec}$ window and add a new
requirement for the reconstructed $W$ to top mass ratio. In
Table~\ref{tab:lowpt_cuts} we quote one working point for illustration
purposes. Two cuts on $\arctan\left(m_{13}/m_{12}\right)$ and
$m_{23}/m_{123}$ are part of the $A$-shaped selection in the original
tagging algorithm, which means all events passing the tagging
algorithm lie inside this $A$-shape. Just as in the ROC analysis of
Section~\ref{sec:default} we now allow for an additional cut on both
of these subjet observables. Finally, we include the best Fox--Wolfram
moments in the four leading jet selections. The cut values quoted in
Table~\ref{tab:lowpt_cuts} define one working point on the ROC curve
for the low-$p_T$ mode. The corresponding efficiencies for this point
are quoted in Table~\ref{tab:lowpt_compare}. Its efficiency over the
$t\bar{t}$ sample with a hadronically decaying top in the same $p_T$
range is significantly increased from 29\% to 35\%, with a slightly
larger fraction of type-2 and type-3 events. In many applications of
the top tagger these low-$p_T$ top decays constitute most of the
signal sample, as shown for the $t\bar{t}H$ production process in
Table~\ref{tab:tth}. Here, the low-$p_T$ improvement might increase
the over-all number of tagged signal events by up to 20\%.\bigskip

To compute an ROC curve for the low-$p_T$ mode we optimize the tagging
parameters only for type-2 tags which did not pass the original
algorithm.  According to Table~\ref{tab:lowpt_compare} optimizing on
all events would give a significant weight to type-3 events, which
would lead to an increased number of tags with very poor top
reconstruction. We therefore compute the ROC curve of the low-$p_T$
tagger by optimizing the tagger for type-2 events in the
$p_{T,\text{fat}} = 150-200$~GeV band. We then compute the
corresponding efficiencies for the entire $t\bar{t}$ sample.  In
Fig.~\ref{fig:lowpt_roc} we compare the performance of the low-$p_T$
mode to the new default tagger shown in
Fig.~\ref{fig:default_roc}. We see first of all that the new mode
makes the slice with $p_{T,\text{fat}} = 150-200$~GeV competitive with
the softest slice in the standard setup. Also the slice with fat jets
within $p_{T,\text{fat}} = 200-250$~GeV benefits from the new mode,
because a large number of newly tagged events with $p_{T,t}^\text{rec}
= 150-200$~GeV appear in that slice. As mentioned above, in the new
version the cuts and tagging criteria can be optimized depending on
the transverse momentum for example of the fat jet, to improve a given
analysis.\bigskip

Before we can make use of the significant improvement documented
above, we need to ensure that for the tops tagged in the low-$p_T$
mode the \textsc{HEPTopTagger} reconstructs the direction and the
4-momentum as well as for the original setup.  In Fig.~\ref{fig:reco}
we see that the dedicated low-$p_T$ tags inside the
\textsc{HEPTopTagger} have almost the same quality of the standard
tags in the same transverse momentum range~\cite{heptop2}.  We show
the geometric distance $\Delta R_t$ between the reconstructed top
momentum and the parton level top momentum as well as the normalized
deviation in transverse momentum, $\Delta p_t/p_t =
(p_{T,t}^\text{rec} - p_{T,t}^\text{true})/p_{T,t}^\text{rec}$.  The
only slight weakness of the low-$p_T$ mode is the depleted region with
excellent angular resolution, \textit{i.e.} $\Delta R_t \lesssim 0.1
\ll R_\text{fat} = 1.8$.  We actually expect that this precision can
be improved by a dedicated calibration procedure, which is beyond the
capabilities of the kind of study we present here.

\section{Outlook}
\label{sec:outlook}

In this top tagging study we defined the new default setup of the
\textsc{HEPTopTagger} for the upcoming 13~TeV run, with a focus on
moderate boosts of $p_{T,t} > 150$~GeV. The entire study is based on
the experimentally accessible semi-leptonic $t\bar{t}$ sample. Our key
results are
\begin{enumerate}
\item For large fat jets with $R_\text{fat} = 1.8$ we define a new
  default setup with reduced shaping of the multi-jet background. In
  addition, we provide a specific high-multiplicity mode for the
  tagger.
\item We give ROC curves which allow us to define the most appropriate
  working points depending on the transverse momentum of the fat
  jet. An optimal working point can be chosen for individual analyses,
  including $p_T$-dependent parameter choices.
\item Including angular correlations after the QCD-inspired selection
  does not improve the tagging performance for $p_{T,\text{fat}} >
  200$~GeV. The same is true for the pruned mass as an additional
  observable. The QCD-inspired tagging algorithm appears to be highly
  efficient in extracting the available information.
\item Similarly, combining the \text{HEPTopTagger} with $N$-Subjettiness
  does not give significant improvement in terms of the ROC curve.
\item We define a low-$p_T$ mode probing tops down to $p_{T,t} =
  150$~GeV, which makes use of angular correlations for tops with only
  part of their decay products captured inside the fat jet. For the
  $t\bar{t}$ sample this leads to a the signal efficiency of 35\% in
  the targeted momentum range.
\end{enumerate}
With those modifications we expect the \textsc{HEPTopTagger} to be
even more useful for a wide range of analyses to come, including
$t\bar{t}H$ searches.\bigskip

\begin{center}
{\bf Acknowledgments}
\end{center}

All of us would like to thank Sebastian Sch\"atzel for his key
contributions which made the \textsc{HEPTopTagger} an experimental
success. 

\newpage



\begin{thebibliography}{99}

\bibitem{review}
 D.~E.~Morrissey, T.~Plehn and T.~M.~P.~Tait,
  Phys.\ Rept.\  {\bf 515}, 1 (2012).

\bibitem{tth}
  T.~Plehn, G.~P.~Salam and M.~Spannowsky,
  Phys.\ Rev.\ Lett.\  {\bf 104}, 111801 (2010).

\bibitem{tth_exp}
 for a nice description of the experimental challenges see 
 J.~Cammin and M.~Schumacher,
  ATL-PHYS-2003-024.

\bibitem{sfitter}
 see \textit{e.g.} 
 T.~Plehn and M.~Rauch,
  Europhys.\ Lett.\  {\bf 100}, 11002 (2012).

\bibitem{2hdm}
 see \textit{e.g.} 
 D.~Lopez-Val, T.~Plehn and M.~Rauch,
  JHEP {\bf 1310}, 134 (2013).

\bibitem{meade}
 P.~Meade and M.~Reece,
  Phys.\ Rev.\  D {\bf 74}, 015010 (2006);
 J.~Ellis, F.~Moortgat, G.~Moortgat-Pick, J.~M.~Smillie and J.~Tattersall,
  Eur.\ Phys.\ J.\  C {\bf 60}, 633 (2009);
 K.~Rolbiecki, J.~Tattersall and G.~Moortgat-Pick,
  Eur.\ Phys.\ J.\ C {\bf 71}, 1517 (2011);
 M.~Perelstein and A.~Weiler,
  JHEP {\bf 0903}, 141 (2009).

\bibitem{heptop}
 T.~Plehn, M.~Spannowsky, M.~Takeuchi, and D.~Zerwas,
  JHEP {\bf 1010}, 078 (2010);
 \url{http://www.thphys.uni-heidelberg.de/~plehn/}

\bibitem{semilep}
 K.~Rehermann and B.~Tweedie,
  JHEP\ {\bf 1103}, 059  (2011);
 T.~Plehn, M.~Spannowsky, M.~Takeuchi,
  JHEP {\bf 1105}, 135 (2011).

\bibitem{early}
 W.~Skiba and D.~Tucker-Smith,
  Phys.\ Rev.\  D {\bf 75}, 115010 (2007);
 B.~Holdom,
  JHEP {\bf 0703}, 063 (2007).
 M.~Gerbush, T.~J.~Khoo, D.~J.~Phalen, A.~Pierce and D.~Tucker-Smith,
  Phys.\ Rev.\  D {\bf 77}, 095003 (2008);
 G.~Brooijmans,
  ATL-PHYS-CONF-2008-008 and ATL-COM-PHYS-2008-001, Feb.~2008.

\bibitem{tt_resonances}
 see \textit{e.g.} 
  K.~Agashe, A.~Belyaev, T.~Krupovnickas, G.~Perez and J.~Virzi,
  Phys.\ Rev.\ D {\bf 77}, 015003 (2008)
 V.~Barger, T.~Han and D.~G.~E.~Walker,
  Phys.\ Rev.\ Lett.\  {\bf 100}, 031801 (2008);
 B.~Lillie, L.~Randall and L.~-T.~Wang,
  JHEP {\bf 0709}, 074 (2007);
 U.~Baur and L.~H.~Orr,
  Phys.\ Rev.\  D {\bf 76}, 094012 (2007);
 U.~Baur and L.~H.~Orr,
  Phys.\ Rev.\  D {\bf 77}, 114001 (2008);
 P.~Fileviez Perez, R.~Gavin, T.~McElmurry and F.~Petriello,
  Phys.\ Rev.\  D {\bf 78}, 115017 (2008).

\bibitem{single_top}
 F.~Kling, T.~Plehn and M.~Takeuchi,
  Phys.\ Rev.\ D {\bf 86}, 094029 (2012).

\bibitem{buckets}
  M.~R.~Buckley, T.~Plehn and M.~Takeuchi,
  JHEP {\bf 1308}, 086 (2013);
 M.~R.~Buckley, T.~Plehn, T.~Schell and M.~Takeuchi,
  arXiv:1310.6034 [hep-ph].
 
\bibitem{tagging_review}
 A.~Abdesselam  {\it et al.},
  Eur.\ Phys.\ J.\ C {\bf 71}, 1661 (2011);
 T.~Plehn and M.~Spannowsky,
  J.\ Phys.\ G {\bf 39}, 083001 (2012);
 A.~Altheimer {\it et al.},
  arXiv:1311.2708 [hep-ex].


\bibitem{seymour}
 M.~H.~Seymour,
  Z.\ Phys.\  C {\bf 62}, 127 (1994);
 J.~M.~Butterworth, B.~E.~Cox and J.~R.~Forshaw,
  Phys.\ Rev.\  D {\bf 65}, 096014 (2002).

\bibitem{bdrs}
  J.~M.~Butterworth, A.~R.~Davison, M.~Rubin and G.~P.~Salam,
  Phys.\ Rev.\ Lett.\  {\bf 100}, 242001 (2008).

\bibitem{hopkins}
 D.~E.~Kaplan, K.~Rehermann, M.~D.~Schwartz and B.~Tweedie,
  Phys.\ Rev.\ Lett.\  {\bf 101}, 142001 (2008).

\bibitem{template}
 L.~G.~Almeida, S.~J.~Lee, G.~Perez, I.~Sung and J.~Virzi,
  Phys.\ Rev.\  D {\bf 79}, 074012 (2009);
 L.~G.~Almeida, S.~J.~Lee, G.~Perez, G.~F.~Sterman, I.~Sung and J.~Virzi,
  Phys.\ Rev.\ D\ {\bf 79}, 074017  (2009);
 L.~G.~Almeida, S.~J.~Lee, G.~Perez, G.~Sterman, I.~Sung,
  Phys.\ Rev.\  {\bf D82}, 054034 (2010).

\bibitem{pruning}
 S.~D.~Ellis, C.~K.~Vermilion and J.~R.~Walsh,
  Phys.\ Rev.\ D\ {\bf 80}, 051501  (2009);
 S.~D.~Ellis, C.~K.~Vermilion and J.~R.~Walsh,
  Phys.\ Rev.\ D\ {\bf 81}, 094023  (2010);
 C.~K.~Vermilion,
  arXiv:1101.1335 [hep-ph].

\bibitem{trimming}
 J.~Thaler, L.~-T.~Wang,
  JHEP {\bf 0807}, 092 (2008);
  D.~Krohn, J.~Thaler and L.~-T.~Wang,
  JHEP\ {\bf 0906}, 059  (2009);
 D.~Krohn, J.~Thaler, L.~-T.~Wang,
  JHEP {\bf 1002}, 084 (2010).

\bibitem{scet}
 J.~Thaler, K.~Van Tilburg,
  JHEP {\bf 1103}, 015 (2011);
 J.~Thaler and K.~Van Tilburg,
  JHEP {\bf 1202}, 093 (2012).

\bibitem{recent}
 A.~Hook, M.~Jankowiak and J.~G.~Wacker,
  JHEP {\bf 1204}, 007 (2012);
 M.~Jankowiak and A.~J.~Larkoski,
  JHEP {\bf 1106}, 057 (2011);
 M.~Jankowiak and A.~J.~Larkoski,
  JHEP {\bf 1204}, 039 (2012);
 D.~E.~Soper and M.~Spannowsky,
  Phys.\ Rev.\ D {\bf 87}, no. 5, 054012 (2013);
 S.~Schaetzel and M.~Spannowsky,
  arXiv:1308.0540 [hep-ph].

\bibitem{atlas}
  ATLAS~Collaboration,
  JHEP {\bf 1301}, 116 (2013);
  ATLAS~Collaboration,
  ATLAS-CONF-2013-084;
  G~Piacquadio 
  CERN-THESIS-2010-027;
  G~Kasieczka,
  PhD thesis, \url{http://www.ub.uni-heidelberg.de/archiv/14941}.

\bibitem{heptop2}
 T.~Plehn, M.~Spannowsky and M.~Takeuchi,
  Phys.\ Rev.\ D {\bf 85}, 034029 (2012).

\bibitem{fwm}
 G.~C.~Fox and S.~Wolfram,
  Phys.\ Rev.\ Lett.\  {\bf 41}, 1581 (1978);
 R.~D.~Field, Y.~Kanev and M.~Tayebnejad,
  Phys.\ Rev.\ D {\bf 55}, 5685 (1997).

\bibitem{fwm_higgs}
 C.~Bernaciak, M.~S.~A.~Buschmann, A.~Butter and T.~Plehn,
  Phys.\ Rev.\ D {\bf 87}, 073014 (2013);
 C.~Bernaciak, B.~Mellado, T.~Plehn, X.~Ruan, and P.~Schichtel,
  arXiv:1311.5891 [hep-ph].

\bibitem{madmax}
 T.~Plehn, P.~Schichtel and D.~Wiegand,
  arXiv:1311.2591 [hep-ph].

\bibitem{alpgen}
 M.~L.~Mangano, M.~Moretti, F.~Piccinini, R.~Pittau and A.~D.~Polosa,
  JHEP {\bf 0307}, 001 (2003).

\bibitem{pythia}
  T.~Sjostrand, S.~Mrenna and P.~Z.~Skands,
  JHEP {\bf 0605}, 026 (2006).

\bibitem{mlm}
 M.~L.~Mangano, M.~Moretti, R.~Pittau,
  Nucl.\ Phys.\  {\bf B632}, 343-362 (2002).

\bibitem{ca_algo}
 Y.~L.~Dokshitzer, G.~D.~Leder, S.~Moretti and B.~R.~Webber,
  JHEP {\bf 9708}, 001 (1997);
 M.~Wobisch and T.~Wengler,
  arXiv:hep-ph/9907280.

\bibitem{pileup}
  ATLAS~Collaboration,
  JHEP {\bf 1309}, 076 (2013).

\bibitem{fastjet}
 M.~Cacciari and G.~P.~Salam,
  Phys.\ Lett.\  B {\bf 641}, 57 (2006);
 M.~Cacciari, G.~P.~Salam and G.~Soyez,
  Eur.\ Phys.\ J.\ C {\bf 72}, 1896 (2012);
 \url{http://fastjet.fr}

\bibitem{jade}
 W.~Bartel {\it et al.}  [JADE Collaboration],
  Z.\ Phys.\  C {\bf 33}, 23 (1986).

\bibitem{tmva}
 A.~H\"ocker {\it et al.},
  PoS ACAT {\bf }, 040 (2007)
  [physics/0703039 [PHYSICS]];
 P.~Speckmayer, A.~H\"ocker, J.~Stelzer and H.~Voss,
  J.\ Phys.\ Conf.\ Ser.\  {\bf 219}, 032057 (2010);
 \url{http://tmva.sourceforge.net}

\bibitem{nsubjettiness} 
 J.~Thaler and K.~Van Tilburg,
  JHEP {\bf 1103}, 015 (2011);
 J.~Thaler and K.~Van Tilburg,
  JHEP {\bf 1202}, 093 (2012);
 I.~W.~Stewart, F.~J.~Tackmann and W.~J.~Waalewijn,
  Phys.\ Rev.\ Lett.\  {\bf 105}, 092002 (2010).

\bibitem{spartyjet} 
  P.~-A.~Delsart, K.~L.~Geerlings, J.~Huston, B.~T.~Martin and C.~K.~Vermilion,
  arXiv:1201.3617 [hep-ex].


\end{thebibliography}
\end{document}